\documentclass[a4paper]{article}

\usepackage[english]{babel}
\usepackage[utf8x]{inputenc}
\usepackage[T1]{fontenc}

\usepackage[a4paper,top=3cm,bottom=2cm,left=3cm,right=3cm,marginparwidth=1.75cm]{geometry}

\usepackage{setspace}
\usepackage{amssymb}

\usepackage{cite}

\usepackage{amsmath}
\usepackage{graphicx}
\usepackage[colorinlistoftodos]{todonotes}
\usepackage[colorlinks=true, allcolors=blue]{hyperref}

\title{Narrow quadrupolar surface lattice resonances and band reversal in vertical metal-insulator-metal gratings}

\author{Xinyu Fang$^{1,2,4,6}$, Lei Xiong$^{1,3,4,6}$, Jianping Shi$^{2}$, Hongwei Ding$^{3}$, and Guangyuan Li$^{1,4,5,*}$}

\date{}
\begin{document}
\maketitle

\begin{spacing}{2.0}

\noindent \large$^1$CAS Key Laboratory of Human-Machine Intelligence-Synergy Systems, Shenzhen Institute of Advanced Technology, Chinese Academy of Sciences, Shenzhen 518055, China

\noindent  $^2$College of Physics and Electronic Technology, Anhui Normal University, Wuhu 241000, China 

\noindent $^3$School of Information Science and Engineering, Yunnan University, Kunming 650500, China

\noindent $^4$Guangdong-Hong Kong-Macao Joint Laboratory of Human-Machine Intelligence-Synergy Systems, Chinese Academy of Sciences, Shenzhen Institute of Advanced Technology, Shenzhen 518055, China

\noindent $^5$Shenzhen College of Advanced Technology, University of Chinese Academy of Sciences, Shenzhen 518055, China

\noindent $^6$These authors contributed equally.

%\noindent $^\dagger$ These authors contributed equally.

\noindent *Corresponding author: gy.li@siat.ac.cn

\end{spacing}

\newpage

\begin{abstract}
We report narrow quadrupolar surface lattice resonances (SLRs) under normal incidence, and the observation, for the first time, of the band reversal effect of SLRs supported by a vertical metal-insulator-metal nanograting, which is embedded in a homogeneous dielectric environment. Simulation results show that under normal incidence, quadrupolar SLR with linewidth of 1~nm and high quality factor of 979 can be excited in the near-infrared regime, and that under oblique incidence, out-of-plane dipolar SLRs of relatively large quality factors ($\geq 150$) can be launched. By varying the incidence angle, the SLR wavelength can be continuously tuned over an extremely broadband range of 750~nm, covering most of the near-infrared regime, and the quality factor decreases exponentially. Remarkably, the resonance lineshape can also be dynamically tuned from an asymmetric Fano-shaped dip to a peak, a dip/peak pair, and a perfect symmetric Lorentzian peak, suggesting the appearance of the band reversal effect. We expect the high-$Q$ SLRs with broadband tunability and tunable lineshapes will find potential applications in enhanced nanoscale light-matter interactions in nanolasers, nonlinear optics and sensing.
\end{abstract}

\newpage

\section{Introduction}
Plasmonic surface lattice resonances (SLRs) supported by periodic arrays of metallic nanostructures have many appealing merits, such as spectrally narrow optical responses, large quality factors, strong field enhancements extended over large volumes, and large wavelength tunability, and thus have emerged as a new platform for enhancing light-matter interactions on the nanoscale  \cite{JPCC2016SchatzSLRreview,MatToday2018OdomSLRreview,ChemRev2018GrigorenkoSLRreview,ACR2019OdomSLRreview}. Over the years, SLRs have found attractive applications in nanolasers \cite{NN2013OdomIPSLRlasing,NM2019OdomSLRlasing,ACSNano2020NorrisSLRlasing}, nonlinear optics \cite{PRL2018HalpinPLR_NL,NL2018KauranenPLR_SHG,NL2019OdomSLR_SHG}, and optical sensing \cite{JPD2016LiuSLRdetection,BioBio2018KabashinSLRsensing,Plasmonics2020_MetasurfaceSensor}.

In order to achieve high quality factors, quadrupolar SLRs have received increasing attention recently. Burrows and Barnes \cite{OE2010Barnes_QuadSLR} made use of large nanoparticles and symmetry-broken conditions (oblique incidences or asymmetric structures) to achieve quadrupolar SLRs. Humphrey {\sl et al.} \cite{ACSP2016Barnes_AsymDimer} employed arrays of asymmetric disc dimers and controlled the relative strength of the dipolar and the quadrupolar SLRs through the degree of asymmetry. Yang {\sl et al.} \cite{PNAS2016Odom_QuadSLR} obtained both dipolar and quadrupolar SLRs in aluminum nanoparticle arrays with diameter larger than 120~nm. However, these quadrupolar SLRs are weakly excited on a dipolar SLR background and have relatively small quality factors. 

In order to tackle this problem, quite recently, the authors \cite{OL2021Li_HVMIM} proposed a horizontal metal-insulator-metal (MIM) nanograting for achieving high-$Q$ out-of-plane quadrupolar SLR without the dipolar SLR background. We showed that the quality factor obtained under oblique incidence can reach up to 1036 in the near-infrared, which is three times of that of the out-of-plane dipolar SLR excited in the vertical MIM nanograting under the same condition. Although these results are encouraging, a question arises: can we excite high-$Q$ quadrupolar SLR without the dipolar SLR background under normal incidence? We are interested in normal incidence since it is the most convenient in experiments.

In this work, we will show that high-$Q$ quadrupolar SLR without dipolar SLR background can be excited in the vertical MIM nanograting under normal incidence. Simulation results will show that the quality factor can reach up to 979 in the near-infrared regime, which is comparable to that of the out-of-plane quadrupolar SLR excited in the horizontal MIM grating under oblique incidence. We will show that, as the incidence angle increases, high-$Q$ out-of-plane dipolar SLRs can be excited, and the SLR wavelength can be tuned over an extremely broadband range, covering 979~nm to 1735~nm. Interestingly, results will show that the SLR quality factor decreases exponentially with the incidence angle but remains above 150. Surprisingly, we will observe, for the first time, the band reversal effect in SLRs, which manifests as the transition of resonance lineshapes from an asymmetric Fano-shape dip to a dip/peak pair, a peak, and a symmetric Lorentzian peak.

\section{Simulation setup}
\begin{figure}[htb]
\centering
\includegraphics[width=70mm]{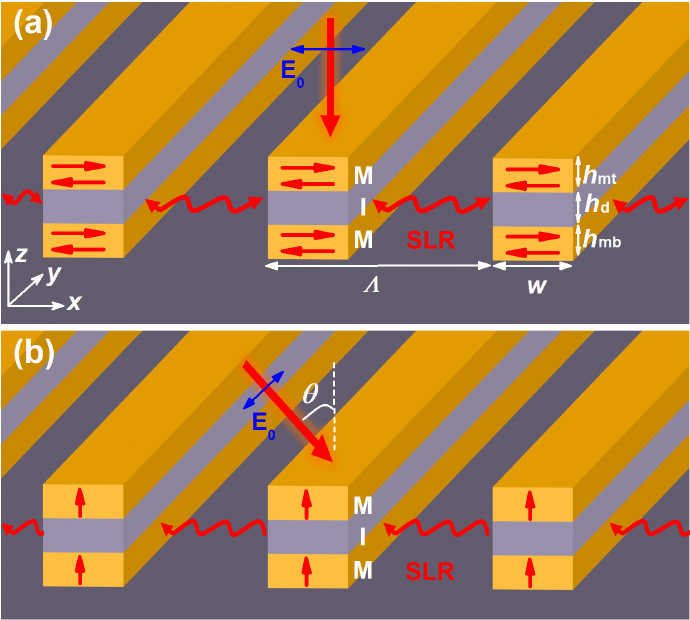}
\caption{Schematics of the vertical MIM nanograting with period $\Lambda$ supports (a) quadrupolar SLR under normal incidence and (b) out-of-plane dipolar SLR under oblique incidence of incidence angle $\theta$ and TM polarization. The nanograting unit cell is composed of vertical gold-silica-gold nanoridges of width $w$ and heights $h_{\rm mt}$, $h_{\rm d}$ and $h_{\rm mb}$ from the top to the bottom. Red arrows in gold nanoridges indicate the directions of the induced electric polarizations.}
\label{fig:schem}
\end{figure}

Figure \ref{fig:schem} illustrates the 1D vertical MIM nanograting  under study, which has period $\Lambda$ in the $x$ direction and is embedded in a homogeneous dielectric environment with refractive index $n_0$. The nanograting unit cell is composed of three gold, silica, and gold nanoridges from the top to the bottom with width of $w$, and heights of $h_{mt}$, $h_{d}$ and $h_{mb}$, respectively. The structure is illuminated by plane wave with unitary electric field amplitude ($|E_0|=1$) at the incidence angle of $\theta$. We restrict ourselves to transverse magnetic (TM) polarization only. 

The zeroth-order reflectance and transmittance spectra, as well as the near-field distributions of the MIM nanograting were simulated with a home-developed package for fully vectorial rigorous coupled-wave analysis (RCWA), which was developed following \cite{JOSAA1995RCWA,JOSAA1996_RCWA1D}. In the simulations, we adopted 301 retained orders, which are tested to be large enough to reach the convergence regime \cite{OL2021Li_HVMIM}. All the calculations were performed with $n_0=n_{\rm d}=1.45$ ($n_0$ is for the silica substrate and the index-matching oil, and $n_{\rm d}$ is for the silica nanoridge), $\Lambda=675$~nm, $w=165$~nm, and $h_{\rm mb}=h_{\rm d}=h_{\rm mt}=100$~nm. Wavelength-dependent permittivities of gold were taken from Johnson and Christy \cite{JC1972NK}.

The MIM nanograting can be fabricated using the state-of-the-art nanofabrication processes. The thin layers of gold, silica and gold films are first deposited on the silica substrate in sequence. A photoresist is spin-coated on the top and patterned using the electron beam lithography. The pattern is then transferred to the gold-silica-gold films using the ion beam etching. Finally, the remaining photoresist is removed, resulting in the vertical MIM nanograting under study.

%%%%%%%%%%%%%%%%%%%%%%%%%%%%%%%%%%%%%%%%%%%%%%%%%%%%%%%%%%%%%%%%%%%%%%%%%%%
\section{Results and discussion}
\subsection{Spectra and near-fields for quadrupolar SLRs}
\begin{figure}[htb]
\centering
\includegraphics[width=\linewidth]{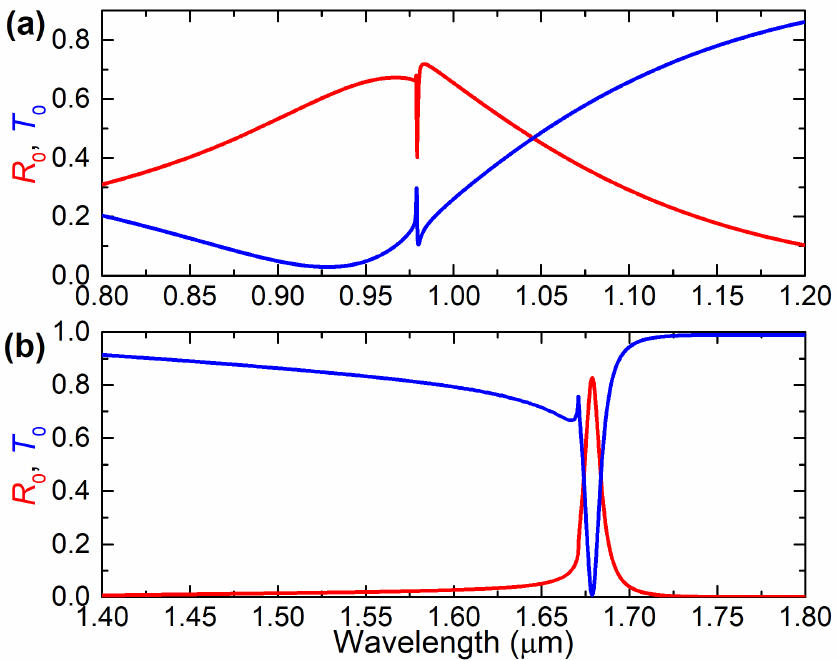}
\caption{Simulated zeroth-order reflectance (red) and transmittance (blue) spectra of the vertical MIM nanograting under (a) normal incidence and (b) oblique incidence of $\theta=45^{\circ}$.}
\label{fig:spectra}
\end{figure}

We first investigate the spectral and near-field responses of the MIM nanograting under normal incidence ($\theta=0^{\circ}$). Figure~\ref{fig:spectra}(a) shows that there exists an extremely narrow Fano-shaped dip (or peak) locating at $\lambda_0=979.4$~nm with a small linewidth of $\Delta \lambda= 1.0$~nm in the reflectance (or transmittance) spectrum. Correspondingly, the quality factor is estimated to be $Q=\lambda_0 /\Delta \lambda = 979.4$. This value is comparable to that ($Q=1036$) of the out-of-plane quadrupolar SLR supported by the horizontal MIM nanograting under oblique incidence \cite{OL2021Li_HVMIM}.

As a reference, we also study the scenario of the oblique incidence of $\theta = 45^{\circ}$, which was done in our previous work \cite{OL2021Li_HVMIM}. Figure~\ref{fig:spectra}(b) shows that the reflectance (or transmittance) spectrum exhibits a relatively broad Fano-shaped peak (or dip) locating at 1678.8~nm. The corresponding linewidth is estimated to be 10.6~nm and the quality factor is calculated to be $Q=158.4$. Compared with our previous work \cite{OL2021Li_HVMIM}, the reflectance and transmittance spectra are similar, and the quality factors are also comparable. 

%%%%%%%%%%%%%%%%%%%%%%%%%%%%%%%%%%%%%%%%%%%
%%%%%%%%%%%%%%%%%%%%%%%%%%%%%%%%%%%%%%%%%%%

\begin{figure}[htp]
\centering
\includegraphics[width=\linewidth]{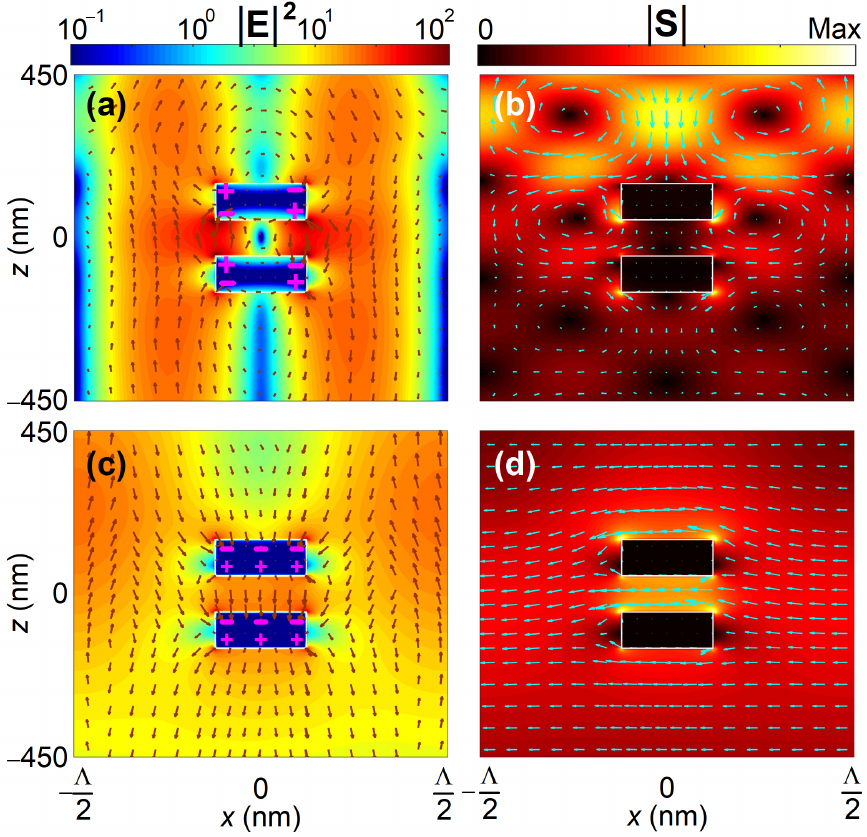}
\caption{(a)(c) Simulated electric field $|E|^2$ and (b)(d) Poynting vector $|S|$ distributions (color for intensity and arrows for directions) at resonance wavelengths of (a)(b) 979.4~nm under normal incidence and (c)(d) 1678.8~nm under the oblique incidence of $\theta=45^\circ$. Symbols ``$+$'' and ``$-$'' in (a)(c) indicate charge distributions. Gold nanoridges are outlined by white rectangles.}
\label{fig:field}
\end{figure}

According to our previous study \cite{OL2021Li_HVMIM}, the asymmetric spectral features in Figure~\ref{fig:spectra} should originate from the Fano-type interference between the localized surface plasmon resonance (LSPR) and the RA, resulting in SLRs. In order to understand the large difference on the spectral linewidth and the quality factor between the two resonances under normal incidence and under the oblique incidence of $\theta=45^\circ$, we plot the simulated near-field electric field $|E|^2$ and Poynting vector $|S|$ maps at the resonance wavelengths of 979.4~nm and 1678.8~nm. Figures~\ref{fig:field}(a)(c) show that the electric fields are greatly enhanced over extended volumes, which are typical characteristics of SLRs, that two quadrupoles are excited in the top and bottom gold nanoridges under normal incidence, and that two out-of-plane dipoles in phase are excited under the oblique incidence. The local electric filed enhancement reaches 210 and 60 for normal incidence and the oblique incidence of $\theta=45^\circ$, respectively. Under normal incidence, the Poynting vector map in Figure~\ref{fig:field}(b) shows circulating energy flows around the gold nanoridges, suggesting standing waves formed by the interference of two counterpropagating surface waves. Under the oblique incidence of $\theta=45^\circ$, however, the energy flux propagates along the $-x$ direction. Therefore, the quadrupolar SLR is excited under normal incidence and the out-of-plane dipolar SLR is launched under the oblique incidence of $\theta=45^\circ$. 

%%%%%%%%%%%%%%%%%%%%%%%%%%%%%%%%%%%%%%%%%%%
%%%%%%%%%%%%%%%%%%%%%%%%%%%%%%%%%%%%%%%%%%%

Note that under normal incidence, the near-field electric field maps in Figure~\ref{fig:field}(a) are distinct from those of MIM nanopillars in our previous work \cite{OE2019_SLRMIM}, where a dipole and a quadrupole are excited in the top and the bottom gold nanoridges, respectively. This difference is induced because here we adopt symmetric dielectric environment, whereas in ref.~\cite{OE2019_SLRMIM} we considered asymmetric environment. Under oblique incidence, the near-field electric field and Poynting vector distributions in Figures~\ref{fig:field}(c)(d) are similar to our previous work\cite{OL2021Li_HVMIM}. The only difference is that, here the SLR wavelength of 1678.8~nm is close to the $-1$ diffraction order of Rayleigh anomaly (RA), because in this work we adopt a much smaller period.

\subsection{Spectral tunability and band reversal effect}
In Figure~\ref{fig:spectra}, we find that, as the incidence angle changes from $\theta=0^\circ$ to $45^\circ$, the SLR wavelength is tuned from 979.4~nm to 1678.8~nm, suggesting extremely broadband tunability. Remarkably, we observe an impressive transition from a Fano-shaped reflectance dip (or a transmittance peak) under normal incidence to a peak (or a dip) under the oblique incidence of $\theta=45^\circ$. This suggests the appearance of the band reversal effect.

We emphasize that for both the quadrupolar SLRs excited in the vertical MIM nanograting under normal incidence and in the horizontal MIM nanograting under oblique incidence \cite{OL2021Li_HVMIM}, there is no dipolar SLR background. The difference in their spectral features is that, the former has a reflectance dip (or transmittance peak), as shown in Figure~\ref{fig:spectra}(a), whereas the latter has a transmittance dip (or reflectance peak) \cite{OL2021Li_HVMIM}, which is similar to Figure~\ref{fig:spectra}(b).

%%%%%%%%%%%%%%%%%%%%%%%%%%%%%%%%%%%%%%%
\begin{figure}[htp]
\centering
\includegraphics[width=\linewidth]{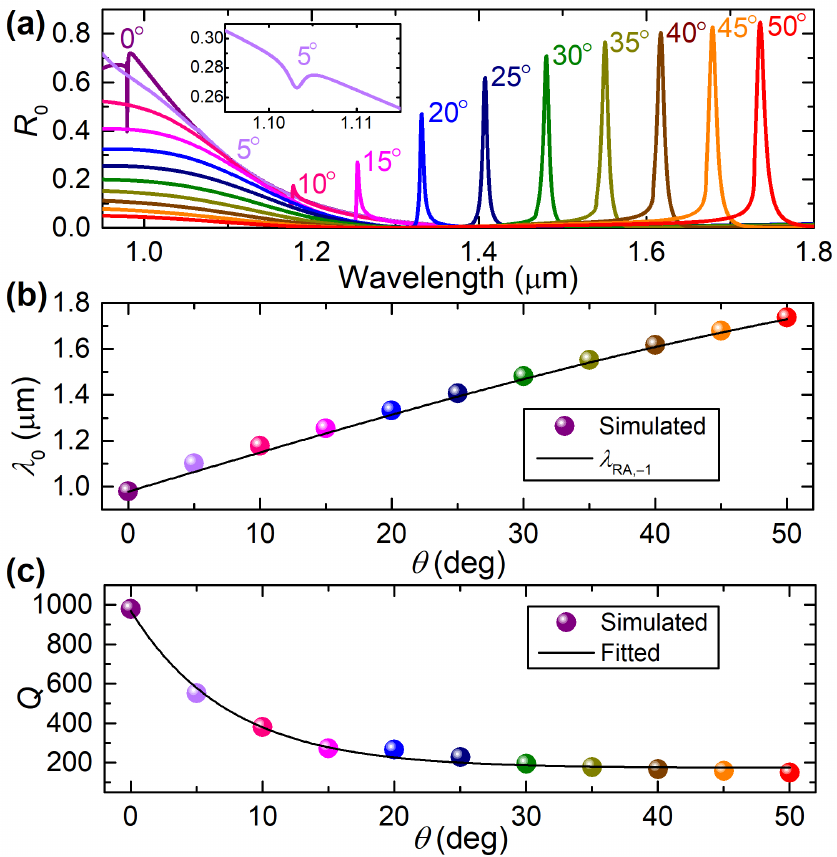}
\caption{(a) Simulated zeroth-order reflectance spectra of the vertical MIM nanograting under incidence of different angles. (b) The corresponding resonance wavelengths and (c) the quality factors as function of the incidence angle. $\lambda_{{\rm RA},m}$ in (b) is given by Equation~(\ref{eq:RA}) and fitted $Q$ is defined by Equation~(\ref{eq:fitQ}).}
\label{fig:RvsAng}
\end{figure}
%%%%%%%%%%%%%%%%%%%%%%%%%%%%%%%%%%%%%%%

To further study the spectral tunability and the band reversal effect, we calculated the reflectance (as shown in Figure~\ref{fig:RvsAng}) and the transmittance (not shown for clarity) spectra for more incidence angles.   Figures~\ref{fig:RvsAng}(a)(b) show that, as the incidence angle increases from $\theta=0^\circ$ to $50^\circ$ in a step of 5$^\circ$, the SLR wavelength increases continuously from $\lambda_0=979.4$~nm to 1735.7~nm. This suggests extremely broadband tunability over 756.3~nm, covering most of the near-infrared regime. These SLR wavelengths generally follow the RA wavelengths of the $m=-1$ order, which are determined by 
%%%%%%%%%%%%%%%%%%%%%%%%%%%%%%%%%%%%%%%
\begin{equation}
\label{eq:RA}
\lambda_{{\rm RA},m}= n_0 \Lambda (\pm 1 - \sin\theta) / m \,.
\end{equation}
%%%%%%%%%%%%%%%%%%%%%%%%%%%%%%%%%%%%%%%

Figure~\ref{fig:RvsAng}(a) also shows that the SLR has an increasing linewidth as the incidence angle increases. We estimate the quality factors for these SLRs that are excited under different incidence angles and plot the results in Figure~\ref{fig:RvsAng}(c). Results show that as the incidence angle increases, the quality factor first decreases dramatically from 979 for $\theta=0^\circ$ to 551 for $\theta=5^\circ$, and then gradually converges to 150 for $\theta=50^\circ$. Interestingly, we further find that the quality factor decreases exponentially following a fitted expressing,
%%%%%%%%%%%%%%%%%%%%%%%%%%%%%%%%%%%%%%%
\begin{equation}
\label{eq:fitQ}
Q = 173 + 796.2 \times \exp{(-0.135\theta)}  \,.
\end{equation}
%%%%%%%%%%%%%%%%%%%%%%%%%%%%%%%%%%%%%%%

Besides the broadband wavelength tunability, which is intrinsic for the SLR, and the exponentially decreasing quality factor as the incidence angle increases, which is interesting, in Figure~\ref{fig:RvsAng}(a) we observe, for the first time, an impressive phenomenon of the band reversal effect in SLRs, which manifest itself as the transition from a Fano-shaped dip to a peak in the reflectance spectra. Although similar effect was reported by Rybin {\sl et al.} in photonic crystals with Fano resonance between continuum Mie scattering and a narrow Bragg band \cite{PRL2009Limonov_BraggReversal}, it has not been reported in the field of SLRs yet. A typical example is the SLR supported by a 1D metal nanoridge grating, for which the transmittance spectra always exhibit a Fano-shaped dip as the incidence angle increases \cite{ACSP2019Odom_1D}.

%%%%%%%%%%%%%%%%%%%%%%%%%%%%%%%%%%%%%%%
\begin{figure}[htp]
\centering
\includegraphics[width=\linewidth]{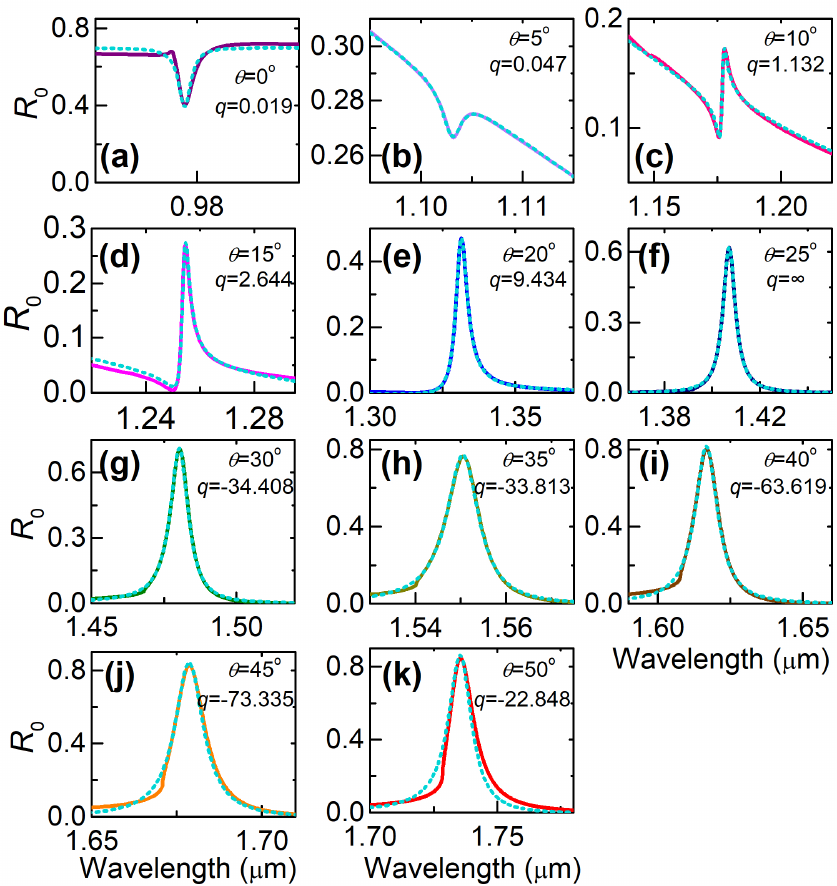}
\caption{Evolution of the resonance lineshape as the incidence angle increases. Solid curves are simulation results taken from Figure~\ref{fig:RvsAng}, and dashed curves are fitted results with equation~\ref{eq:fitq} using Fano parameters $q$ labelled on the top right.}
\label{fig:LineShape}
\end{figure}
%%%%%%%%%%%%%%%%%%%%%%%%%%%%%%%%%%%%%%%

In order to focus on the evolution of the resonance lineshape, we replot the reflectance spectra around the SLR wavelengths in Figure~\ref{fig:LineShape}. It is clear that the resonance lineshape evolves from a Fano-shaped dip for $\theta=0^\circ$ and $5^\circ$, to a dip/peak pair with a distinct asymmetric Fano lineshape for $\theta=10^\circ$ and $15^\circ$, then a reflectance peak with symmetric Lorentzian lineshape for $\theta=25^\circ$, and finally a peak with quasi-Lorentzian lineshape for larger incidence angles. 

The evolution of the resonance lineshape can be quantified using the Fano parameter $q$, which can be extracted by fitting the reflectance spectra with\cite{RMP2010Kivshar_FanoRev,NP2017Kivshar_FanoRev}
%%%%%%%%%%%%%%%%%%%%%%%%%%%%%%%%%%%%%%%
\begin{equation}
\label{eq:fitq}
R_0 = B + A \frac{ (q \gamma + \lambda - \lambda_0)^2}{(\lambda - \lambda_0)^2+\gamma^2}  \,.
\end{equation}
%%%%%%%%%%%%%%%%%%%%%%%%%%%%%%%%%%%%%%%
Here the fitting parameters include the background spectrum $B$, the coefficient $A$, the Fano parameter $q$, and the nonradiative linewidth $\gamma$. 

%%%%%%%%%%%%%%%%%%%%%%%%%%%%%%%%%%%%%%%
\begin{figure}[htp]
\centering
\includegraphics[width=\linewidth]{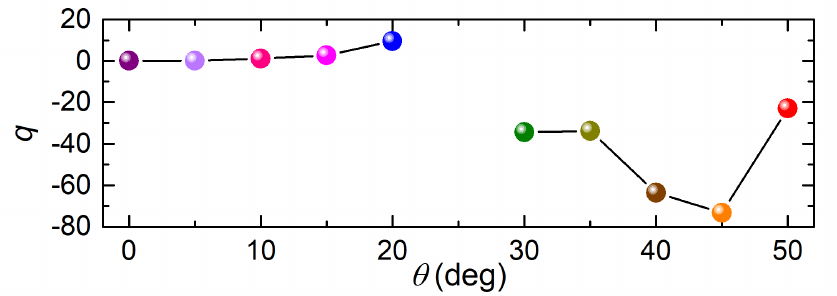}
\caption{Fano parameter $q$ as a function of the incidence angle.}
\label{fig:qvsAng}
\end{figure}
%%%%%%%%%%%%%%%%%%%%%%%%%%%%%%%%%%%%%%%

The extracted Fano parameters for the different incidence angles are labelled in Figure~\ref{fig:LineShape}, and the evolution as a function of the incidence angle is depicted in Figure~\ref{fig:qvsAng}. We find that $q=0.019$ and 0.047 for $\theta=0^\circ$ and $5^\circ$, respectively, corresponding Fano-shaped dips in the reflectance spectra. $q=1.132$ and 2.644 for $\theta=10^\circ$, $15^\circ$, respectively. These correspond to a distinct asymmetric Fano dip/peak lineshape. For $\theta=20^\circ$, $q=9.434$, since the reflectance peak is less asymmetric. Particularly, for $\theta=25^\circ$, $q\rightarrow \infty$, since the reflectance peak has a perfect Lorentzian lineshape. As the incidence angle further increases, the Fano parameters $q$ have negative values with large magnitudes, which correspond to less asymmetric Fano lineshapes. Therefore, we show that by varying the incidence angle, the SLR lineshape can be dynamically tuned ranging from a Fano-shaped dip, a Fano-shaped dip/peak pair, a Fano-shaped peak, to a perfect Lorentzian peak. This suggests that the vertical MIM nanograting can act as a platform for realizing Lorentzian and Fano resonance lineshapes, which is similar to the microring resonator/bus-waveguide structure \cite{NanoP2019Gan_FanoLorentzian}.

\section{Conclusions}
In summary, we have shown that narrow quadrupolar SLR with high quality factor of 979 in the near-infrared regime can be excited under normal incidence in the vertical MIM nanograting. Under oblique incidence, out-of-plane dipolar SLRs can be excited. We have found that these SLRs have extremely broadband tunability (over 756~nm) while maintaining relatively high quality factors (above 150). Interestingly, we have found that the quality factor decreases exponentially as the incidence angle increases. Remarkably, we have observed, for the first time, the band reversal effect in the field of SLRs, which manifests as the lineshape evolution from a Fano-shaped dip, to a dip/peak pair of distinct Fano lineshape, a Fano-shaped peak, and a peak with perfect Lorentzian lineshape. This makes the vertical MIM nanograting an attractive platform for realizing dynamically tunable Fano and Lorentzian resonance lineshapes. We expect the SLRs with high quality factors, extremely broadband tunability, and tunable lineshapes will find promising applications in light-matter interactions on the nanoscale.

%%%%%%%%%%%%%%%%%%%%%%%%%%%%%%%%%%%

\section*{Acknowledgments}
This work was funded by the State Key Laboratory of Advanced Optical Communication Systems and Networks, China (2019GZKF2).
%\end{acknowledgments}

\bibliographystyle{unsrt}
\bibliography{sample}

\begin{thebibliography}{10}

\bibitem{JPCC2016SchatzSLRreview}
M.~B. Ross, C.~A. Mirkin, and G.~C. Schatz.
\newblock Optical properties of one-, two-, and three-dimensional arrays of
  plasmonic nanostructures.
\newblock {\em J. Phys. Chem. C}, 120:816--830, 2016.

\bibitem{MatToday2018OdomSLRreview}
W.~Wang, M.~Ramezani, A.~I. V{\"a}kev{\"a}inen, P.~T{\"o}rm{\"a}, J.~G. Rivas,
  and T.~W. Odom.
\newblock The rich photonic world of plasmonic nanoparticle arrays.
\newblock {\em Mater. Today}, 21:303--314, 2018.

\bibitem{ChemRev2018GrigorenkoSLRreview}
V.~G. Kravets, A.~V. Kabashin, W.~L. Barnes, and A.~N. Grigorenko.
\newblock Plasmonic surface lattice resonances: A review of properties and
  applications.
\newblock {\em Chem. Rev.}, 118:5912--5951, 2018.

\bibitem{ACR2019OdomSLRreview}
D.~Wang, J.~Guan, J.~Hu, M.~R. Bourgeois, and T.~W. Odom.
\newblock Manipulating light-matter interactions in plasmonic nanoparticle
  lattices.
\newblock {\em Acc. Chem Res.}, 52:2997--3007, 2019.

\bibitem{NN2013OdomIPSLRlasing}
W.~Zhou, M.~Dridi, J.~Y. Suh, C.~H. Kim, D.~T. Co, M.~R. Wasielewski, G.~C.
  Schatz, and T.~W. Odom.
\newblock Lasing action in strongly coupled plasmonic nanocavity arrays.
\newblock {\em Nat. Nanotechnol.}, 8:506--511, 2013.

\bibitem{NM2019OdomSLRlasing}
A.~Fernandez-Bravo et~al.
\newblock Ultralow-threshold, continuous-wave upconverting lasing from
  subwavelength plasmons.
\newblock {\em Nat. Mat.}, 18:1172--1176, 2019.

\bibitem{ACSNano2020NorrisSLRlasing}
Jan~M. Winkler, Max~J. Ruckriegel, Henar Rojo, Robert~C. Keitel, Eva~De Leo,
  Freddy~T. Rabouw, and David~J. Norris.
\newblock Dual-wavelength lasing in quantum-dot plasmonic lattice lasers.
\newblock {\em ACS Nano}, 14:5223--5232, 2020.

\bibitem{PRL2018HalpinPLR_NL}
Mohammad Ramezani, Quynh Le-Van, and Alexei Halpin.
\newblock Nonlinear emission of molecular ensembles strongly coupled to
  plasmonic lattices with structural imperfections.
\newblock {\em Phys. Rev. Lett.}, 121:243904, 2018.

\bibitem{NL2018KauranenPLR_SHG}
R.~Czaplicki, A.~Kiviniemi, M.~J. Huttunen, X.~Zang, T.~Stolt, I.~Vartiainen,
  J.~Butet, M.~Kuittinen, O.~J.~F. Martin, and M.~Kauranen.
\newblock Less is more: Enhancement of second-harmonic generation from
  metasurfaces by reduced nanoparticle density.
\newblock {\em Nano Lett.}, 18:7709--7714, 2018.

\bibitem{NL2019OdomSLR_SHG}
D.~C. Hooper, C.~Kuppe, D.~Wang, W.~Wang, J.~Guan, T.~W. Odom, and V.~K. Valev.
\newblock Second harmonic spectroscopy of surface lattice resonances.
\newblock {\em Nano Lett.}, 19:165--172, 2019.

\bibitem{JPD2016LiuSLRdetection}
Zhengqi Liu, Guiqiang Liu, Guolan Fu, Xiaoshan Liu, Zhenping Huang, and Gang
  Gu.
\newblock All-metal meta-surfaces for narrowband light absorption and high
  performance sensing.
\newblock {\em J. Phys. D: Appl. Phys.}, 49:445104, 2016.

\bibitem{BioBio2018KabashinSLRsensing}
A.~Danilov, G.~Tselikov, F.~Wu, V.~G. Kravets, I.~Ozerov, F.~Bedu, A.~N.
  Grigorenko, and A.~V. Kabashin.
\newblock Ultra-narrow surface lattice resonances in plasmonic metamaterial
  arrays for biosensing applications.
\newblock {\em Biosens. Bioelectron.}, 104:102--112, 2018.

\bibitem{Plasmonics2020_MetasurfaceSensor}
Mohammad~Reza Rakhshani.
\newblock Tunable and sensitive refractive index sensors by plasmonic absorbers
  with circular arrays of nanorods and nanotubes for detecting cancerous cells.
\newblock {\em Plasmonics}, 15:2071--2080, 2020.

\bibitem{OE2010Barnes_QuadSLR}
Christopher~P. Burrows and William~L. Barnes.
\newblock Large spectral extinction due to overlap of dipolar and quadrupolar
  plasmonic modes of metallic nanoparticles in arrays.
\newblock {\em Opt. Express}, 18:3187--3198, 2010.

\bibitem{ACSP2016Barnes_AsymDimer}
Alastair~D. Humphrey, Nina Meinzer, Timothy~A. Starkey, and William~L. Barnes.
\newblock Surface lattice resonances in plasmonic arrays of asymmetric disc
  dimers.
\newblock {\em ACS Photon.}, 3:634--639, 2016.

\bibitem{PNAS2016Odom_QuadSLR}
Ankun Yang, Alexander~J Hryn, Marc~R Bourgeois, Won-Kyu Lee, Jingtian Hu,
  George~C Schatz, and Teri~W Odom.
\newblock Programmable and reversible plasmon mode engineering.
\newblock {\em Proc. Natl. Acad. Sci. U. S. A.}, 113(50):14201--14206, 2016.

\bibitem{OL2021Li_HVMIM}
Xinyu Fang, Lei Xiong, Jianping Shi, and Guangyuan Li.
\newblock High-$q$ quadrupolar plasmonic lattice resonances in horizontal
  metal-insulator-metal gratings.
\newblock {\em Opt. Lett.}, 46(7):1546--1549, 2021.

\bibitem{JOSAA1995RCWA}
M.~G. Moharam, Drew~A. Pommet, Eric~B. Grann, and T.~K. Gaylord.
\newblock Stable implementation of the rigorous coupled-wave analysis for
  surface-relief gratings: Enhanced transmittance matrix approach.
\newblock {\em J. Opt. Soc. Am. A}, 12(5):1077--1086, 1995.

\bibitem{JOSAA1996_RCWA1D}
Philippe Lalanne and G.~Michael Morris.
\newblock Highly improved convergence of the coupled-wave method for tm
  polarization.
\newblock {\em J. Opt. Soc. Am. A}, 13:779--784, 1996.

\bibitem{JC1972NK}
P.~B. Johnson and R.~W. Christy.
\newblock Optical constants of the noble metals.
\newblock {\em Phys. Rev. B}, 6:4370--4379, 1972.

\bibitem{OE2019_SLRMIM}
Xiuhua Yang, Gongli Xiao, Yuanfu Lu, and Guangyuan Li.
\newblock Narrow plasmonic surface lattice resonances with preference to
  asymmetric dielectric environment.
\newblock {\em Opt. Express}, 27:25384--25394, 2019.

\bibitem{PRL2009Limonov_BraggReversal}
M.V. Rybin, A.~B. Khanikaev, M.~Inoue, K.~B. Samusev, M.~J. Steel, G.~Yushin,
  and M.~F. Limonov.
\newblock Fano resonance between mie and bragg scattering in photonic crystals.
\newblock {\em Phys. Rev. Lett.}, 103:023901, 2009.

\bibitem{ACSP2019Odom_1D}
Y.~Hua, A.~K. Fumani, and T.~W. Odom.
\newblock Tunable lattice plasmon resonances in {1D} nanogratings.
\newblock {\em ACS Photon.}, 6(2):322--326, 2019.

\bibitem{RMP2010Kivshar_FanoRev}
Andrey~E. Miroshnichenko, Sergej Flach, and Yuri~S. Kivshar.
\newblock Fano resonances in nanoscale structures.
\newblock {\em Rev. Mod. Phys.}, 82:2257--2298, 2010.

\bibitem{NP2017Kivshar_FanoRev}
Mikhail~F. Limonov, Mikhail~V. Rybin, Alexander~N. Poddubny, and Yuri~S.
  Kivshar.
\newblock Fano resonances in photonics.
\newblock {\em Nat. Photon.}, 11:543--554, 2017.

\bibitem{NanoP2019Gan_FanoLorentzian}
Linpeng Gu, Hanlin Fang, Juntao Li, Liang Fang, Soo~Jin Chua, Jianlin Zhao, and
  Xuetao Gan.
\newblock A compact structure for realizing lorentzian, fano, and
  electromagnetically induced transparency resonance lineshapes in a microring
  resonator.
\newblock {\em Nanophotonics}, 8:841--848, 2019.

\end{thebibliography}

\end{document}